\documentclass[doublecol,preprint,longbibliography,superscriptaddress,nofootinbib]{epl2}
\usepackage[utf8]{inputenc}
\usepackage{graphicx}
\usepackage{amssymb}
\usepackage{textcomp}
\usepackage{amsmath}
\usepackage{tabularx}
\usepackage{bm}
\usepackage{times}
\usepackage{color}
\usepackage{slashed}
\usepackage{multirow}
\usepackage{verbatim}
\usepackage{cancel}
\usepackage{subfigure}
\usepackage{ulem}

\usepackage[colorlinks=true, pdfstartview=FitV, linkcolor=blue, citecolor=blue, urlcolor=blue]{hyperref}
\allowdisplaybreaks[4]

\def\lsim{\mathrel{\raise.3ex\hbox{$<$\kern-.75em\lower1ex\hbox{$\sim$}}}}
\def\gsim{\mathrel{\raise.3ex\hbox{$>$\kern-.75em\lower1ex\hbox{$\sim$}}}}

\definecolor{orange}{rgb}{1,0.5,0}

\title{Sensitivity of two-mode SRF cavity to generic electromagnetism of sub-$\mu$eV dark matter}

\author{Chang-Jie Dai\inst{1} \and Tong Li\inst{1} \and Rui-Jia Zhang\inst{1}}

\institute{
  \inst{1} School of Physics, Nankai University, Tianjin 300071, China
}

\abstract{
The low-mass dark matter (DM) such as axion or wavelike scalar is a plausible DM candidate. Recently, the possible non-standard couplings of low-mass DM has drawn much attention. In this work we investigate the detection of electromagnetic couplings in a few benchmark models of low-mass DM. For illustration, we consider the generic axion electrodynamics including CP violating coupling as well as the newly proposed axion electromagnetodynamics.
The superconducting radio frequency (SRF) cavity with two modes has more advantages than the traditional cavity approach with static background field. We utilize the two-mode SRF cavity to probe the generic couplings of low-mass DM with frequency lower than GHz.
We show the sensitivity of the SRF cavity to the axion couplings in the above frameworks.}

\begin{document}

\maketitle

%%%%%%%%%%%%%%%%%%%%%%%%%%%%%%%%
\section{Introduction}
\label{sec:Intro}
%%%%%%%%%%%%%%%%%%%%%%%%%%%%%%%%

The low-mass dark matter (DM) such as axion, axion-like particle (ALP) and light scalar DM is a plausible candidate for DM with feeble interaction, .
The Peccei-Quinn (PQ) mechanism provides a solution of the strong CP problem by introducing a pseudo-Goldstone axion field $a$ after the spontaneous breaking of $U(1)_{\rm PQ}$ symmetry~\cite{Peccei:1977hh,Peccei:1977ur,Weinberg:1977ma,Wilczek:1977pj,Baluni:1978rf,Crewther:1979pi,Kim:1979if,Shifman:1979if,Dine:1981rt,Zhitnitsky:1980tq,Baker:2006ts,Pendlebury:2015lrz}. The low-mass scalar DM $\phi$ is also induced by a coherently oscillating field and can explain the DM observation.
Recently, a lot of works studying the non-standard couplings of low-mass DM appeared in the literature.
In general, there would be the CP-violating couplings between low-mass DM and electromagnetic fields~\cite{Gorghetto:2021luj,Luo:2023cxo,DiLuzio:2023cuk}
\begin{eqnarray}
aF^{\mu\nu}F_{\mu\nu}~~{\rm or}~~\phi F^{\mu\nu}\tilde{F}_{\mu\nu}\;.
\end{eqnarray}
They are expected to result in more interesting phenomenologies in particle physics and cosmology.

On the other hand, there exist many recent studies discussing the possible modification of standard axion electrodynamics inspired by the framework of quantum electromagnetodynamics (QEMD)~\cite{Schwinger:1966nj,Zwanziger:1968rs,Zwanziger:1970hk} in both theory~\cite{Sokolov:2022fvs,Sokolov:2023pos,Heidenreich:2023pbi,Li:2023zcp} and phenomenology~\cite{Li:2022oel,Tobar:2022rko,McAllister:2022ibe,Li:2023kfh,Li:2023aow,Tobar:2023rga,Patkos:2023lof}. The QEMD theory inherently introduces both electric and magnetic charges, and is able to build the connection to the Witten effect~\cite{Witten:1979ey}. Ref.~\cite{Sokolov:2022fvs} constructed a more generic axion-photon Lagrangian in the low-energy axion effective field theory (EFT). New anomalous axion-photon interactions and couplings arise assuming the existence of heavy PQ-charged fermions with electric
and magnetic charges. As a result of the above generic axion-photon Lagrangian, the conventional axion Maxwell equations~\cite{Sikivie:1983ip} have been further modified and some consequent new detection strategies of axion have been studied in recent years~\cite{Li:2022oel,Tobar:2022rko,McAllister:2022ibe,Li:2023kfh}.

In this work, we investigate the sensitivity of superconducting radio frequency (SRF) cavity with two modes to the above general interactions of low-mass DM.
This approach was proposed to detect resonant axion using the cavity with frequency difference between two modes~\cite{Sikivie:2010fa,Goryachev:2018vjt,Thomson:2019aht,Berlin:2019ahk,Lasenby:2019prg,Lasenby:2019hfz,Bogorad:2019pbu,Berlin:2020vrk,Gao:2020anb,Salnikov:2020urr,Gorghetto:2021luj,Berlin:2022hfx}. Instead of typical searches using
static magnetic fields, the initial ``pump mode'' is induced by a magnetic field oscillating in time with a high frequency $\omega_0\gg m_a$. The ``signal mode'' is then tuned to have a similar frequency $\omega_1\simeq \omega_0 + m_a$.
The axion is upconverted to
the frequency of the readout mode. This is the upconversion technique. The high frequency saturates a large quality factor $Q$ and thus the SRF cavity becomes a perfect environment for this approach. Meanwhile, this axion-induced frequency conversion in SRF cavity is sensitive to the much lower mass of axion DM than the typical LC regime.
We will adjust the setups of SRF cavity for the above general interactions of low-mass DM and show the sensitivity of adjusted SRF cavity to different couplings in sub-$\mu$eV mass region.

%%%%%%%%%%%%%%%%%%%%%%%%%%%%%%%%%%%%%%%%%
\section{The generic electromagnetism of low-mass DM}
\label{sec:genericaxion}
%%%%%%%%%%%%%%%%%%%%%%%%%%%%%%%%%%%%%%%%%

In this section we describe a few benchmark frameworks with generic electromagnetic interactions.

%%%%%%%%%%%%%%%%%%%%%%%%%%%%
\subsection{The generic electrodynamics of axion or scalar low-mass DM}
\label{sec:ULDMaxion}
%%%%%%%%%%%%%%%%%%%%%%%%%%%%

The generic electromagnetic interactions of axion low-mass DM including CP violating coupling are
\begin{eqnarray}
\mathcal{L}_{\rm axion} \supset -{\frac{1}{4}} g_{a\gamma\gamma} a F_{\mu\nu} F^{d~\mu\nu} - {\frac{1}{4}} g_{a\gamma\gamma}^{\cancel{\rm CP}} a F_{\mu\nu} F^{\mu\nu} \;,
\end{eqnarray}
where $F_{\mu\nu}=\partial_\mu A_\nu - \partial_\nu A_\mu$ with $A_\mu$ being the four-potential of the $U(1)_{\rm EM}$ group, and $F^d_{\mu\nu}\equiv\tilde{F}_{\mu\nu}=\epsilon_{\mu\nu\alpha\beta}F^{\alpha\beta}/2$ with $\epsilon_{0123}=+1$ as the Hodge dual of tensor $F_{\mu\nu}$.
Our results for axion can be directly applied to scalar DM by using the substitution $a\to \phi\;,~~g_{a\gamma\gamma} \to  g_{\phi\gamma\gamma}^{\cancel{\rm CP}}\;,~~g_{a\gamma\gamma}^{\cancel{\rm CP}} \to  g_{\phi\gamma\gamma}$.
After applying the Euler-Lagrange equation of motion for the potential $A_\mu$, one obtains
the following equations with axion field
\begin{eqnarray}
&&\partial_\mu F^{\mu\nu} +g_{a\gamma\gamma} \partial_\mu (a F^{d~\mu\nu}) + g_{a\gamma\gamma}^{\cancel{\rm CP}} \partial_\mu (a F^{\mu\nu}) = J_{e}^\nu\;,\\
&&\partial_\mu F^{d~\mu\nu} =0 \;,
\end{eqnarray}
where $J_{e}$ is the electromagnetic current. By ignoring the term proportional to the small quantity $|g_{a\gamma\gamma}^{\cancel{\rm CP}} a| \ll 1$~\footnote{The axion field can be given by $|a|\sim \sqrt{2\rho_{\rm DM}}/m_a$ with $\rho_{\rm DM}=0.4~{\rm GeV}~{\rm cm}^{-3}$ being the local DM density. Suppose $g_{a\gamma\gamma}^{\cancel{\rm CP}}\sim {1\over f_a}$ and the axion mass-PQ scale relation, we have $|g_{a\gamma\gamma}^{\cancel{\rm CP}} a|\sim 10^{-18}$.}, the new Maxwell equations for axion in terms of electric and magnetic fields are given by~\footnote{We use symbols ``$\mathbb{B}$'' and ``$\mathbb{E}$'' to denote magnetic field and electric field, respectively.}
\begin{eqnarray}
&&\vec{\nabla}\cdot \vec{\mathbb{E}} = \rho_e + g_{a\gamma\gamma} \vec{\mathbb{B}}\cdot \vec{\nabla} a - g_{a\gamma\gamma}^{\cancel{\rm CP}} \vec{\mathbb{E}}\cdot \vec{\nabla} a \;,\\
&&\vec{\nabla}\cdot \vec{\mathbb{B}} = 0 \;,\\
&&\vec{\nabla}\times \vec{\mathbb{E}}+{\partial \vec{\mathbb{B}}\over \partial t}=0\;,
\label{eq:Max3}\\
&&\vec{\nabla}\times \vec{\mathbb{B}}-{\partial \vec{\mathbb{E}}\over \partial t}=\vec{j}_{e}+g_{a\gamma\gamma}(\vec{\mathbb{E}} \times \vec{\nabla} a - {\frac{\partial a}{\partial t}} \vec{\mathbb{B}})\nonumber \\
&&
+ g_{a\gamma\gamma}^{\cancel{\rm CP}} (\vec{\mathbb{B}} \times \vec{\nabla} a + {\frac{\partial a}{\partial t}} \vec{\mathbb{E}}) \;,
\end{eqnarray}
where $\rho_e$ and $\vec{j}_e$ denote the electric charge and current, respectively.

%%%%%%%%%%%%%%%%%%%%%%%%%%%%%%%%%%%%%%%%%
\subsection{The electromagnetodynamics of axion}
\label{sec:QEMDaxion}
%%%%%%%%%%%%%%%%%%%%%%%%%%%%%%%%%%%%%%%%%

In the QEMD theory, the gauge group of QED is replaced with $U(1)_{\rm E}\times U(1)_{\rm M}$ which inherently introduces both electric and magnetic charges. The photon is instead described by two four-potentials $A^\mu$ and $B^\mu$. The non-trivial form of equal-time canonical commutation
relations between them guarantees the preservation of the right degrees of
freedom of physical photon~\cite{Zwanziger:1970hk}.
The complete Lagrangian for the generic interactions between axion and four-potentials based on QEMD is~\footnote{We define $(X \wedge Y)^{\mu \nu}\equiv X^\mu Y^\nu - X^\nu Y^\mu$ for any four-vectors $X$ and $Y$.}~\cite{Sokolov:2022fvs}
\begin{eqnarray}
&&\mathcal{L}= {1\over 2n^2} \{[n\cdot (\partial \wedge B)]\cdot [n\cdot (\partial \wedge A)^d] \nonumber \\
&&-[n\cdot (\partial \wedge A)]\cdot [n\cdot (\partial \wedge B)^d] - [n\cdot (\partial\wedge A)]^2 - [n\cdot (\partial\wedge B)]^2\} \nonumber \\
&&-{1\over 4} g_{aAA}~a~{\rm tr}[(\partial \wedge A)(\partial \wedge A)^d] - {1\over 4} g_{aBB}~a~{\rm tr}[(\partial \wedge B)(\partial \wedge B)^d]\nonumber \\
&&- {1\over 2} g_{aAB}~a~{\rm tr}[(\partial \wedge A)(\partial \wedge B)^d] -J_e \cdot A - J_m\cdot B +\mathcal{L}_G \;,
\end{eqnarray}
where $J_e$ and $J_m$ denote the electric and magnetic currents, respectively, and $\mathcal{L}_G$ is a gauge-fixing term. The $g_{aAA}$ coupling is equivalent to the conventional $g_{a\gamma\gamma}$ coupling.

After applying the Euler-Lagrange equation of motion for the two potentials, one obtains
the following axion modified Maxwell equations~\cite{Sokolov:2022fvs}
\begin{eqnarray}
&&\partial_\mu F^{\mu\nu} -g_{aAA} \partial_\mu a F^{d~\mu\nu} + g_{aAB} \partial_\mu a F^{\mu\nu} = J_e^\nu\;,  \\
&&\partial_\mu F^{d~\mu\nu} +g_{aBB} \partial_\mu a F^{\mu\nu} - g_{aAB} \partial_\mu a F^{d~\mu\nu} = J_m^\nu\;,
\end{eqnarray}
where the term responsible for Witten effect is omitted.
The new Maxwell equations in terms of electric and magnetic fields are then given by
\begin{eqnarray}
&&\vec{\nabla}\cdot \vec{\mathbb{E}} = \rho_e + g_{aAA} \vec{\mathbb{B}}\cdot \vec{\nabla} a - g_{aAB} \vec{\mathbb{E}}\cdot \vec{\nabla} a \;,\\
&&\vec{\nabla}\cdot \vec{\mathbb{B}} = \rho_m -g_{aBB} \vec{\mathbb{E}}\cdot \vec{\nabla} a + g_{aAB} \vec{\mathbb{B}}\cdot \vec{\nabla} a \;,\\
&&\vec{\nabla}\times \vec{\mathbb{E}}+{\partial \vec{\mathbb{B}}\over \partial t}=\vec{j}_m-g_{aBB}(\vec{\mathbb{B}} \times \vec{\nabla} a + {\partial a\over \partial t} \vec{\mathbb{E}})\nonumber \\
&&
- g_{aAB} (\vec{\mathbb{E}} \times \vec{\nabla} a - {\partial a\over \partial t} \vec{\mathbb{B}})\;,
\label{eq:QEMDmax3}\\
&&\vec{\nabla}\times \vec{\mathbb{B}}-{\partial \vec{\mathbb{E}}\over \partial t}=\vec{j}_e+g_{aAA}(\vec{\mathbb{E}} \times \vec{\nabla} a - {\partial a\over \partial t} \vec{\mathbb{B}})\nonumber \\
&&
+ g_{aAB} (\vec{\mathbb{B}} \times \vec{\nabla} a + {\partial a\over \partial t} \vec{\mathbb{E}})\;,
\label{eq:QEMDmax4}
\end{eqnarray}
where the magnetic charge $\rho_m$ and current $\vec{j}_m$ will be ignored below as there is no observed magnetic monopole.

%%%%%%%%%%%%%%%%%%%%%%%%%%%%%%%%
\section{The signal power for generic couplings of low-mass DM and noise sources in SRF cavity}
\label{sec:SRF}
%%%%%%%%%%%%%%%%%%%%%%%%%%%%%%%%%

In this section we follow Ref.~\cite{Berlin:2019ahk} to calculate the signal power and extend the approach for ordinary axion to the above generic axion couplings. We also summarize the noise sources in the SRF cavity.

%%%%%%
\subsection{The signal power of generic axion electrodynamics}
%%%%%%

For the new Maxwell equations of axion in Sec.~\ref{sec:ULDMaxion}, we first apply the curl operation to the Faraday's law Eq.~(\ref{eq:Max3})
\begin{eqnarray}
\Vec{\nabla}(\Vec{\nabla}\cdot\Vec{\mathbb{E}})-\Vec{\nabla}^2 \Vec{\mathbb{E}}+\partial_t (\nabla\times\Vec{\mathbb{B}})=0\;.
\end{eqnarray}
The Gauss's law and the Amp\`{e}re's circuital law are then inserted. Under the approximation with $\Vec{\nabla}a \approx 0$ and no free electric charge and current, we obtain
\begin{eqnarray}
(\nabla^2-\partial^2_t)
\vec{\mathbb{E}}=-g_{a\gamma\gamma}\partial_t(\Vec{\mathbb{B}}\partial_t a)+g_{a\gamma\gamma}^{\cancel{\rm CP}}\partial_t(\Vec{\mathbb{E}}\partial_t a)\;.
\end{eqnarray}
It turns out that we can impose background magnetic field or electric field to detect coupling $g_{a\gamma\gamma}$ or $g_{a\gamma\gamma}^{\cancel{\rm CP}}$, respectively.
We then expand the electric and magnetic fields by the vacuum modes of cavity
\begin{eqnarray}
\Vec{\mathbb{E}}(t,\vec{r})=\sum_{n} e_n(t)\Tilde{\Vec{\mathbb{E}}}_n(\Vec{r})\;,~~
\Vec{\mathbb{B}}(t,\vec{r})=\sum_{n} b_n(t)\Tilde{\Vec{\mathbb{B}}}_n(\Vec{r})\;.
\end{eqnarray}
After taking the Fourier transformation for the above equation of motion, the cavity's electric field can be obtained in the presence of background magnetic field for $g_{a\gamma\gamma}$ coupling
\begin{eqnarray}
&&\sum_{n}\Big(\omega^2-\omega^2_n-i\frac{\omega \omega_n}{Q_n}\Big)\Vec{\mathbb{E}}_n(\omega)\nonumber \\
&&=-g_{a\gamma\gamma}\int dt ~e^{-i\omega t}\partial_t(\Vec{\mathbb{B}} \partial_t a)\;,
\end{eqnarray}
where $\Vec{\mathbb{E}}_n(\omega)=\int dt~ e^{-i\omega t} e_n(t)  \Tilde{\Vec{\mathbb{E}}}_n(\Vec{r}
)=e_n(\omega) \Tilde{\Vec{\mathbb{E}}}_n(\Vec{r}
)$, a dissipating term $i\omega \omega_n/Q_n$ is introduced by hand on the left-handed side, $Q_n$ is the quality factor for each mode, and we define $\omega^2-\omega^2_n-i\frac{\omega \omega_n}{Q_n}=k_n$.
The Fourier transform of the time part of the vacuum cavity mode $n=1$ then becomes
\begin{eqnarray}
&&e_1(\omega)=\int dt e^{-i\omega t} e_1(t)\nonumber \\
&&=g_{a\gamma\gamma} \frac{-\omega}{k_1(\omega)} \int {d\omega'\over 2\pi} (\omega-\omega') b_0(\omega') a(\omega-\omega') \frac{\int_{V} \Tilde{\Vec{\mathbb{E}}}_1^*\cdot\Tilde{\Vec{\mathbb{B}}}_0}{\int_{V} \left| \Tilde{\Vec{\mathbb{E}}}_1(\Vec{r})\right|^2}\;,
\end{eqnarray}
where the background magnetic field is given by the pump mode $\Vec{\mathbb{B}}\approx \Vec{\mathbb{B}}_0=b_0(t)\Tilde{\Vec{\mathbb{B}}}_0(\Vec{r})$.

Note that the time average of a function $f$ can be written in terms of the power spectral density (PSD) $S_f(\omega)$
\begin{eqnarray}
&&\langle f(t)^2\rangle={1\over (2\pi)^2} \int d\omega S_f(\omega)\;,\\
&&S_f(\omega)\delta(\omega-\omega')=\langle f(\omega)f^\ast(\omega')\rangle\;,
\label{eq:Sf}
\end{eqnarray}
where $\langle \cdots \rangle$ denotes the time average of function $f$.
The PSD of signal power in the SRF cavity becomes
\begin{eqnarray}
S_{\rm sig}(\omega)&=&{\omega_1\over Q_1} (g_{a\gamma\gamma}\eta_{\mathbb{E}_1\mathbb{B}_0}\mathbb{B}_0)^2 V  \frac{\omega^2}{(\omega^2-\omega_1^2)^2+(\omega\omega_1/Q_1)^2}\nonumber \\
&\times&\int \frac{d\omega'}{(2\pi)^2} S_{b_0}(\omega')~(\omega'-\omega)^2 S_a(\omega'-\omega)\;,
\end{eqnarray}
where $\eta_{\mathbb{E}_1\mathbb{B}_0}=\frac{\int_V \Tilde{\Vec{\mathbb{E}}}_1^*\cdot \Tilde{\Vec{\mathbb{B}}}_0}
{\sqrt{\int_V|\Tilde{\Vec{\mathbb{E}}}_1|^2}\sqrt{\int_V|\Tilde{\Vec{\mathbb{B}}}_0|^2}}$ and $\mathbb{B}_0=\sqrt{\frac{1}{V}\int_V |\Tilde{\Vec{\mathbb{B}}}_0|^2}$.
When taking a monochromatic pump source $S_{b_0}(\omega)=\pi^2 [\delta(\omega-\omega_0)+\delta(\omega+\omega_0)]$, we have
\begin{eqnarray}
&&S_{\rm sig}(\omega)={\omega_1\over Q_1} (g_{a\gamma\gamma}\eta_{\mathbb{E}_1\mathbb{B}_0}\mathbb{B}_0)^2 V  \frac{\omega^2}{(\omega^2-\omega_1^2)^2+(\omega\omega_1/Q_1)^2}\nonumber \\
&&\times\Big[(\omega-\omega_0)^2 S_a(\omega-\omega_0)+(\omega+\omega_0)^2 S_a(\omega+\omega_0)\Big]\;.
\label{eq:Ssig}
\end{eqnarray}
In the practical calculations, we use the following Maxwellian velocity distribution for $S_a(\omega)$ in Eq.~(\ref{eq:Ssig})~\cite{Berlin:2020vrk}
\begin{align}
S_a(\omega)=\Theta(|\omega|-m_a)\frac{2\pi^2 \rho_{\rm DM}}{m_a^3 v_a^2}e^{-(|\omega|-m_a)/(m_a v_a^2)}\;,
\end{align}
where $\Theta$ is the Heaviside step function and the dispersion velocity is $v_a\approx 9\times 10^{-4}\approx 1/\sqrt{Q_a}$~\cite{Berlin:2020vrk}.
For the CP violating coupling, we just need to make the following substitution
\begin{eqnarray}
g_{a\gamma\gamma}\to g_{a\gamma\gamma}^{\cancel{\rm CP}}\;,~~\eta_{\mathbb{E}_1 \mathbb{B}_0}\to \eta_{\mathbb{E}_1 \mathbb{E}_0}\;,~~\mathbb{B}_0\to \mathbb{E}_0\;.
\end{eqnarray}

Next we should make a choice of the transverse electromagnetic modes (TM and TE) for the overlap factor $\eta_{\mathbb{E}_1 \mathbb{B}_0}$ or $\eta_{\mathbb{E}_1 \mathbb{E}_0}$. For a cylindrical cavity with $(r,\varphi,z)$ coordinates, the electric and magnetic field components of
a TM mode are defined as
\begin{eqnarray}
&&\mathbb{B}_z^{\rm TM}=0\;,~~\mathbb{E}_z^{\rm TM}=\psi(r,\varphi)\cos\Big({p\pi z\over L}\Big)\;,\\
&&\label{transverse B}
\vec{\mathbb{B}}_T^{\rm TM}={i\omega_{mnp}\over \gamma_{mn}^2} \cos\Big({p\pi z\over L}\Big) \hat{z}\times \nabla_{T}\psi(r,\varphi)\;,\\
&&\vec{\mathbb{E}}_T^{\rm TM}=-{p\pi\over L\gamma_{mn}^2} \sin\Big({p\pi z\over L}\Big) \nabla_{T}\psi(r,\varphi)\;,
\end{eqnarray}
where $L$ denotes the height of the cylinder.
The function $\psi(r,\varphi)$ becomes
\begin{eqnarray}
    \psi(r,\varphi) = E_0 J_m(\gamma_{mn}r)e^{im\varphi}\;,
\end{eqnarray}
where $\gamma_{mn}=x_{mn}/R$ with $x_{mn}$ being the $n$th zero of the $m$th order Bessel function $J_m(x)$, $R$ is the cylinder radius and the frequency of the ${\rm TM}_{mnp}$ mode is given by $\omega_{mnp}^2=\gamma_{mn}^2+(p\pi/L)^2$.
The electric and magnetic field components of
a TE mode are defined as
\begin{eqnarray}
&&\mathbb{B}_z^{\rm TE}=\phi(r,\varphi)\sin\Big({p\pi z\over L}\Big)\;,~~\mathbb{E}_z^{\rm TE}=0\;,\\
&&\vec{\mathbb{B}}_T^{\rm TE}={p\pi\over L\gamma_{mn}^{\prime 2}} \cos\Big({p\pi z\over L}\Big) \nabla_{T}\phi(r,\varphi)\;,\\
&&\label{transverse E}
\vec{\mathbb{E}}_T^{\rm TE}=-{i\omega_{mnp}\over \gamma_{mn}^{\prime 2}} \sin\Big({p\pi z\over L}\Big) \hat{z}\times \nabla_{T}\phi(r,\varphi)\;.
\end{eqnarray}
The function $\phi(r,\varphi)$ becomes
\begin{eqnarray}
    \phi(r,\varphi) = B_0 J_m(\gamma^\prime_{mn}r)e^{im\varphi}\;,
\end{eqnarray}
where $\gamma^\prime_{mn}=x^\prime_{mn}/R$ with $x^\prime_{mn}$ being the $n$th root of the $m$th order Bessel function $J^\prime_m(x)$, and the frequency of the ${\rm TE}_{mnp}$ mode is given by $\omega_{mnp}^2=\gamma_{mn}^{\prime 2}+(p\pi/L)^2$.
We find that the $\eta$ factor equals to zero only in three cases
\begin{eqnarray}
0&=&\int_V \Vec{\mathbb{E}}^{\rm TM}\cdot \Vec{\mathbb{B}}^{\rm TM}\propto J_m(\gamma_{m_1 n_1}R)J_m(\gamma_{m_0 n_0}R)\nonumber\\
&=&\int_V \Vec{\mathbb{B}}^{\rm TE}\cdot \Vec{\mathbb{B}}^{\rm TM}\propto J_m(\gamma^\prime_{m_{1(0)} n_{1(0)}}R)J_m(\gamma_{m_{0(1)} n_{0(1)}}R)\nonumber\\
&=&\int_V \Vec{\mathbb{E}}^{\rm TM}\cdot \Vec{\mathbb{E}}^{\rm TE}\propto J_m(\gamma_{m_{1(0)} n_{1(0)}}R)J_m(\gamma^\prime_{m_{0(1)} n_{0(1)}}R) \;.\nonumber\\
\end{eqnarray}
This selection rule limits the choices of the transverse electromagnetic modes for individual low-mass DM coupling.
We show all non-zero modes for both $g_{a\gamma\gamma}$ and $g_{a\gamma\gamma}^{\cancel{\rm CP}}$ couplings in Table~\ref{tab:ULDMmodes1}. The cases with magnetic (electric) field in the pump mode are above (below) the double-line. The modes with check mark are those we choose in our calculations.

\begin{table}[htb!]
\centering
%\resizebox{\textwidth}{!}{
\begin{tabular}{c|c|c}
\hline
     couplings & $g_{a\gamma\gamma}$ & $g_{a\gamma\gamma}^{\cancel{\rm CP}}$  \\
%    \hline
    \hline
    modes & $\vec{\mathbb{E}}_1^{{\rm TE}_{021}}\cdot \vec{\mathbb{B}}_0^{{\rm TM}_{030}}$ ($\checkmark$) & \\
     & $\vec{\mathbb{E}}_1^{\rm TM}\cdot \vec{\mathbb{B}}_0^{\rm TE}$ &  \\
     & $\vec{\mathbb{E}}_1^{\rm TE}\cdot \vec{\mathbb{B}}_0^{\rm TE}$ & \\
    \hline
    \hline
    modes &  & $\vec{\mathbb{E}}_1^{{\rm TM}_{131}}\cdot \vec{\mathbb{E}}_0^{{\rm TM}_{121}}$ ($\checkmark$)\\
     &  & $\vec{\mathbb{E}}_1^{\rm TE}\cdot \vec{\mathbb{E}}_0^{\rm TE}$ \\
     \hline
\end{tabular}
%}
\caption{The selected pump mode and signal mode for both $g_{a\gamma\gamma}$ and $g_{a\gamma\gamma}^{\cancel{\rm CP}}$ couplings in generic axion electrodynamics.
}
\label{tab:ULDMmodes1}
\end{table}

Given the above selected modes, the factor $\eta_{\mathbb{E}_1\mathbb{B}_0}$ for $g_{a\gamma\gamma}$ becomes
\begin{eqnarray}
\eta_{\mathbb{E}_1\mathbb{B}_0}&=&\frac{\int_V \Tilde{\Vec{\mathbb{E}}}_1^\ast\cdot \Tilde{\Vec{\mathbb{B}}}_0}
{\sqrt{\int_V|\Tilde{\Vec{\mathbb{E}}}_1|^2}\sqrt{\int_V|\Tilde{\Vec{\mathbb{B}}}_0|^2}}\simeq 0.70\;.
\end{eqnarray}
Note that $\eta_{\mathbb{E}_1\mathbb{B}_0}$ does not depend on the height of cavity $L$ and we take $R\approx 0.4$ m in the integrals.
For $g_{a\gamma\gamma}^{\cancel{\rm CP}}$ coupling, note that $\eta_{\mathbb{E}_1\mathbb{E}_0}$ is relevant to both the length $ L$ and radius $ R$ of cavity. When fixing $R \approx 5$ m and $L\approx 0.01$ m, we obtain $\eta_{\mathbb{E}_1\mathbb{E}_0}\approx 0.13$.

%%%%%%%%%%%%%%%%%%%%%%%%
\subsection{The signal power of axion electromagnetodynamics}
%%%%%%%%%%%%%%%%%%%%%%%%

For the new Maxwell equations of axion based on QEMD in Sec.~\ref{sec:QEMDaxion}, we also apply the curl operation to the Faraday's law Eq.~(\ref{eq:QEMDmax3}) and the Amp\`{e}re's law Eq.~(\ref{eq:QEMDmax4}).
We obtain
\begin{align}
\label{QEMD wave vibration 1}
(\nabla^2-\partial^2_t)
\vec{\mathbb{B}}&=-g_{aBB}\partial_t(\Vec{\mathbb{E}}\partial_t a)+g_{aAB}\partial_t(\Vec{\mathbb{B}}\partial_t a)\;,\\
\label{QEMD wave vibration 2}
(\nabla^2-\partial^2_t)\vec{\mathbb{E}}&=-g_{aAA}\partial_t(\Vec{\mathbb{B}}\partial_t a)+g_{aAB}\partial_t(\Vec{\mathbb{E}}\partial_t a) \;.
\end{align}
After expanding the electromagnetic fields and taking the Fourier transformation, the cavity's fields become
\begin{align}
\sum_{n}(\omega^2-\omega^2_n-i\frac{\omega \omega_n}{Q_n})\Vec{\mathbb{B}}_n(\omega)&=-g_{aBB}\int dt e^{-i\omega t}\partial_t(\Vec{\mathbb{E}} \partial_t a) \notag \\
&+g_{aAB}\int dt e^{-i\omega t}\partial_t(\Vec{\mathbb{B}} \partial_t a)
\label{QEMD fourier 1}\;,\\
\sum_{n}(\omega^2-\omega^2_n-i\frac{\omega \omega_n}{Q_n})\Vec{\mathbb{E}}_n(\omega)&=-g_{aAA}\int dt e^{-i\omega t}\partial_t(\Vec{\mathbb{B}} \partial_t a) \notag \\
&+g_{aAB}\int dt e^{-i\omega t}\partial_t(\Vec{\mathbb{E}} \partial_t a)\label{QEMD fourier 2}\;.
\end{align}
It turns out that, given a fixed background field, the presence of $g_{aAB}$ and $g_{aAA}$ (or $g_{aBB}$) couplings can induce both electric and magnetic fields in the cavity. If we want to select the signal from either $g_{aAB}$ or $g_{aAA}$ (or $g_{aBB}$) coupling, we have to properly choose specific signal mode and pump mode according to the selection rules of TM and TE modes.
In Table~\ref{tab:ULDMmodes}, we show the selected modes for $g_{aAA}$, $g_{aAB}$ and $g_{aBB}$ couplings in QEMD axion model. One can detect the signal from only one coupling in these experimental setups.

\begin{table}[htbp!]
%\centering
%\scriptsize
\renewcommand{\arraystretch}{1.2}
\hspace*{-0.03\columnwidth}
\resizebox{1.0\columnwidth}{!}{
\begin{tabular}{c|c|c|c}
\hline
     couplings & $g_{aAA}$ & $g_{aAB}$ & $g_{aBB}$ \\
%    \hline
    \hline
    modes & $\vec{\mathbb{E}}_1^{{\rm TE}_{021}}\cdot \vec{\mathbb{B}}_0^{{\rm TM}_{030}}$ ($\checkmark$) &  & \\
     & $\vec{\mathbb{E}}_1^{\rm TM}\cdot \vec{\mathbb{B}}_0^{\rm TE}$ &  $\vec{\mathbb{B}}_1^{{\rm TM}_{131}}\cdot \vec{\mathbb{B}}_0^{{\rm TM}_{121}}$ ($\checkmark$)  & \\
    \hline
    \hline
    modes &  & $\vec{\mathbb{E}}_1^{{\rm TM}_{131}}\cdot \vec{\mathbb{E}}_0^{{\rm TM}_{121}}$ ($\checkmark$)& $\vec{\mathbb{B}}_1^{{\rm TM}_{030}}\cdot \vec{\mathbb{E}}_0^{{\rm TE}_{021}}$ ($\checkmark$)  \\
     &  &  & $\vec{\mathbb{B}}_1^{\rm TE}\cdot \vec{\mathbb{E}}_0^{\rm TM}$ \\
     \hline
\end{tabular}
}
\caption{The selected pump mode and signal mode for $g_{aAA}$, $g_{aAB}$ and $g_{aBB}$ couplings in QEMD model.
}
\label{tab:ULDMmodes}
\end{table}

The overlap factor for $g_{aAA}$ is exactly the same as that for $g_{a\gamma\gamma}$. The $g_{aAB}$ coupling can induce the cavity signals of electric TM mode and magnetic TM mode with electric TM background and magnetic TM background, respectively. The case of electric fields is the same as that for $g_{a\gamma\gamma}^{\cancel{\rm CP}}$ coupling. For the case of magnetic fields, the overlap factor is given by
\begin{eqnarray}
\eta_{\mathbb{B}_1 \mathbb{B}_0}&=&
\frac{\int_V \Tilde{\Vec{\mathbb{B}}}_1^\ast\cdot \Tilde{\Vec{\mathbb{B}}}_0}
{\sqrt{\int_V|\Tilde{\Vec{\mathbb{B}}}_1|^2}\sqrt{\int_V|\Tilde{\Vec{\mathbb{B}}}_0|^2}}\;.
\end{eqnarray}
In this case, the $p$ integer should be the same for both pump mode and signal mode. We choose $p_0=p_1=1$. To ensure $\int_V \Tilde{\Vec{\mathbb{B}}}_1^\ast\cdot \Tilde{\Vec{\mathbb{B}}}_0\neq 0$, we choose $m_0=m_1=1$. Note that $\eta_{\mathbb{B}_1 \mathbb{B}_0}$ does not depend on cavity height $L$ and agrees with the above result of $\eta_{\mathbb{E}_1 \mathbb{E}_0}$ when the $L$ dependent terms vanish. We take $R\approx 5 $ m and also obtain $\eta_{\mathbb{B}_1 \mathbb{B}_0}\approx 0.13$ which meets the result of the above $\eta_{\mathbb{E}_1 \mathbb{E}_0}$ with a small $L\approx 0.01$ m.
For $g_{aBB}$, the $\eta$ factor becomes
\begin{eqnarray}
\eta_{\mathbb{B}_1 \mathbb{E}_0}&=&
\frac{\int_V \Tilde{\Vec{\mathbb{B}}}_1^\ast\cdot \Tilde{\Vec{\mathbb{E}}}_0}
{\sqrt{\int_V|\Tilde{\Vec{\mathbb{B}}}_1|^2}\sqrt{\int_V|\Tilde{\Vec{\mathbb{E}}}_0|^2}}\simeq 0.70\;.
\end{eqnarray}
In this case, the sum of two $p$ integers should be odd and thus we choose $p_0=1$ and $p_1=0$. We also take $R\approx 0.4$ m in the above integrals.
%%%%%%%%%%%%%%%%%%%%%%%%%

%%%%%%
\subsection{Summarized noise sources}
%%%%%%

In this section we summarize the noise sources described in Refs.~\cite{Berlin:2019ahk,Berlin:2020vrk}.

\underline{\textbf{The mechanical vibration noise}} The mechanical noise such as the thermal excitations of the cavity or external vibrations can induce a static shift in the cavity mode frequencies. This shift of resonant frequencies leads to a modification of cavity modes and one can solve the modified equation of motion to find the mechanical vibration noise PSD. The PSD of mechanical noise is given by~\cite{Berlin:2019ahk,Berlin:2020vrk}
\begin{eqnarray}
S_{\rm mech}(\omega)&=&\sum_{n=0,1}S^{(n)}_{\rm mech}(\omega)\;.
\end{eqnarray}
The frequency of the mechanical normal modes of the cavity is defined as $\omega_m$.
Direct measurements found a forest of mechanical resonances $\omega_m^i$ above $\omega_{\rm min}=1$ kHz, separated in frequency by $\mathcal{O}(100)$ Hz~\cite{Bernard:2001kp}.
For each scanned $m_a$, if $m_a$ is close to an $\omega_m^i$, the mechanical noise is more severe due to the resonance between mechanical vibration and electromagnetic mode. If $|m_a-\omega_m^i|\approx 50$ Hz which is half of the average distance between two adjacent $\omega_m^i$, the mechanical noise is minimized. We take the average choice in the practical calculation: $\omega_m=m_a+25$ Hz for $m_a>\omega_{\rm min}$. For $m_a<\omega_{\rm min}$, there are no nearby resonances and we take $\omega_m=\omega_{\rm min}$. A peak-like structure happens at $m_a\simeq \omega_{\rm min}-25~{\rm Hz}=975~{\rm Hz}$.

\underline{\textbf{The thermal noise}}
The cavity wall emits radio waves and induces the so-called thermal noise. The PSD of thermal noise is given by~\cite{Berlin:2019ahk}
\begin{align}
S_{\rm th}(\omega)=\frac{Q_1}{Q_{\rm int}}\frac{4\pi T (\omega\omega_1/Q_1)^2}{(\omega^2-\omega_1^2)^2+(\omega\omega_1/Q_1)^2}\;,
\end{align}
where $T$ is the temperature which the cavity is cooled at. We take the cavity temperature as $T=1.8$ K.

\underline{\textbf{The amplifier noise}} An amplifier is set to readout the signal from the resonant cavity. A flat PSD of the amplifier noise power is $S_{\rm ql}(\omega)=4\pi \omega_1$.
This amplifier noise is given under the assumption of single photon quantum limit.

\underline{\textbf{The oscillator phase noise}}
For background magnetic field, the pump mode is supposed to be excited by an external oscillator. The PSD of the oscillator phase noise is given by~\cite{Berlin:2019ahk}
\begin{align}
S_{\rm phase}(\omega)&\simeq\frac{1}{2}\epsilon^2_{1d}S_{\varphi}(\omega-\omega_0)\nonumber\\
&\frac{(\omega\omega_1/Q_1)^2}{(\omega^2-\omega_1^2)^2+(\omega\omega_1/Q_1)^2}\frac{\omega_0 Q_1}{\omega_1 Q_0}P_{\rm in}\;,
\end{align}
where the spectrum of a commercial oscillator is approximately taken to be a flat form $S_{\varphi}(m_a)$ and follows a fitted result in Ref.~\cite{Berlin:2019ahk}
\begin{align}
S_{\varphi}(\omega)=\sum^3_{n=0}b_n\omega^{-n}
\end{align}
with $b_0=10^{-16}~{\rm Hz}^{-1}$, $b_1=10^{-9}$, $b_2=10^{-6}~{\rm Hz}$, and $b_3=10^{-5}~{\rm Hz}^{2}$.

%%%%%%%%%%%%%%%%%%%%%%%%%%%%%%%%
\section{Sensitivity of two-mode SRF cavity to electromagnetic interactions of low-mass DM}
\label{sec:Cavity}
%%%%%%%%%%%%%%%%%%%%%%%%%%%%%%%%%

We are able to calculate the axion-induced signal power in terms of the PSD in Eq.~(\ref{eq:Ssig})
\begin{eqnarray}
P_{\rm sig}= {1\over (2\pi)^2} \int d\omega S_{\rm sig}(\omega)\;.
\end{eqnarray}
The frequencies of cavity modes are related to the length and radius of the cavity. In this two-mode SRF cavity experiment, the signal mode and the pump mode need to be nearly degenerate. The degeneracy requires the ratio of cavity length and radius to satisfy
\begin{eqnarray}
\Big(\frac{L}{R}\Big)^2=\frac{\pi^2(p_1^2-p_0^2)}{x^2_{mn_0}-x'^2_{mn_1}}\;.
\end{eqnarray}

The total noise becomes
\begin{align}
S_{\rm noise}(\omega)&=&S_{\rm ql}(\omega)+\frac{Q_1}{Q_{\rm cpl}}(S_{\rm th}(\omega)+S_{\rm phase}(\omega)\nonumber \\
&&+S^{(1)}_{\rm mech}(\omega))+\frac{Q_0}{Q_{\rm cpl}}S^{(0)}_{\rm mech}(\omega)\;,
\end{align}
where an additional factor ${Q_1}/{Q_{\rm cpl}}$ is introduced to describe the signal power transmitted from cavity to readout waveguide.
The last term describes the mechanical noise from pump mode with frequency $\omega_0$. One adds a pre-factor ${Q_0}/{Q_{\rm cpl}}$ to describe the coupling between readout and pump mode.
Then, the signal-to-noise ratio (SNR) is given by
\begin{align}
({\rm SNR})^2\simeq t_{\rm int}\int^{\infty}_{0} d\omega \Big(\frac{S_{\rm sig}(\omega)}{S_{\rm noise}(\omega)}\Big)^2\;,
\end{align}
where $t_{\rm int}\simeq t_e~{\Delta \omega_{sc}}/{m_a}$ with an e-fold time $t_e=10^7~s$ and $\Delta \omega_{sc} \simeq {\rm max}(m_a/Q_a,\omega_1/Q_1)$ being single scan width.
We require ${\rm SNR} > 1$ to show the sensitivity of the two-mode SRF cavity to the generic axion couplings. For $g_{a\gamma\gamma}$, $g_{aAA}$ and $g_{aBB}$ couplings, we assume the cavity column as $V=1~{\rm m}^3$ and the factor of $\eta\simeq 1$. For $g_{a\gamma\gamma}^{\cancel{\rm CP}}$ and $g_{aAB}$, we instead fix $R\approx 5$ m and vary the length $L$ from 0.01 m to 10 m.

\begin{figure}[htb!]
\centering
\includegraphics[scale=0.15]{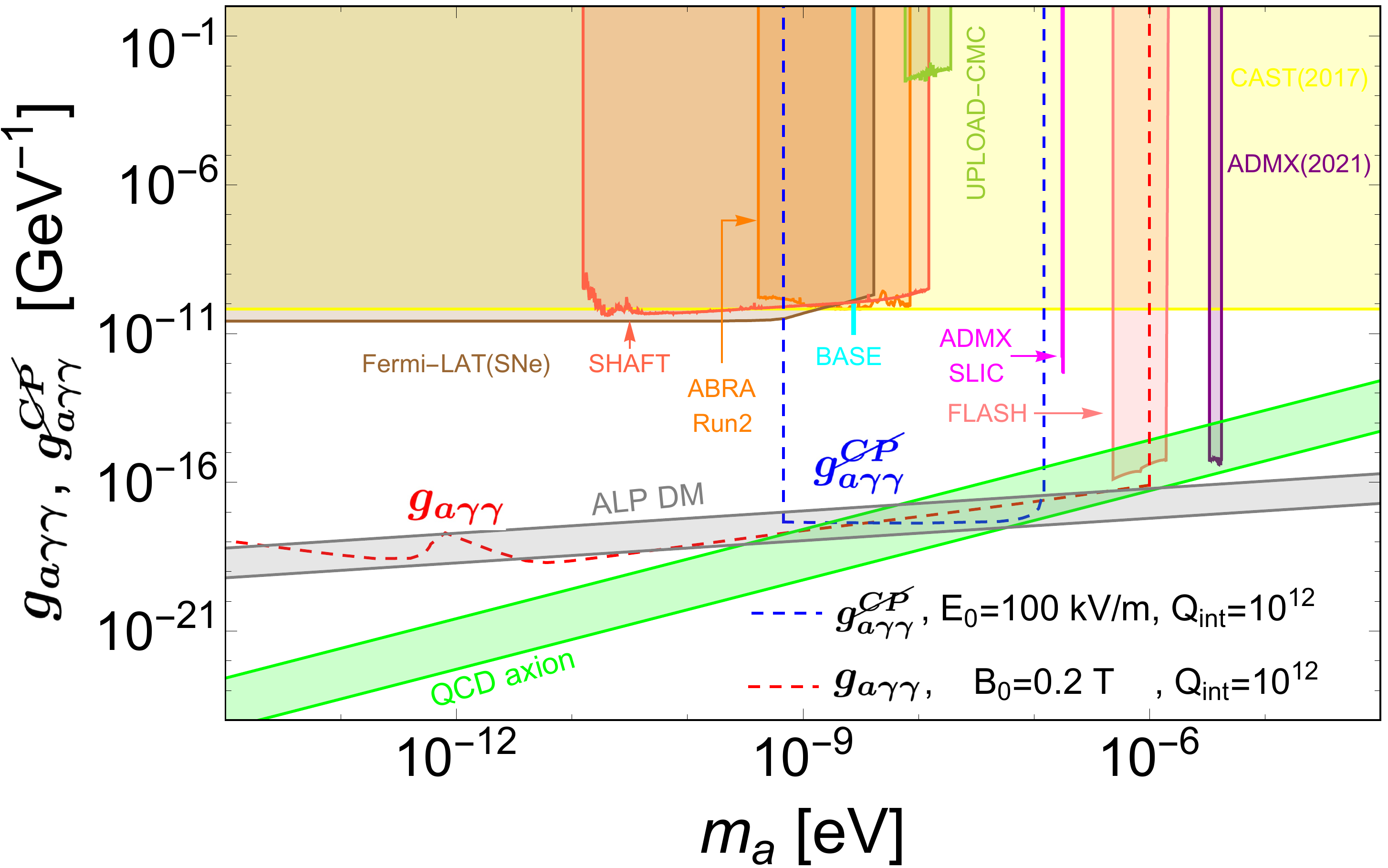}
\caption{
The expected sensitivity bounds of $g_{a\gamma\gamma}$ with $B_0=0.2~{\rm T}$ (red dashed line) and $g_{a\gamma\gamma}^{\cancel{\rm CP}}$ with $E_0=100~{\rm kV/m}$ (blue dashed line) as a function of $m_a$. The theoretical predictions of $g_{a\gamma\gamma}$ from QCD axion (green region)~\cite{DiLuzio:2020wdo} and ALP DM for the standard cosmology with the initial misalignment angle of $\theta_0\in [0.1,1]$ (gray region)~\cite{Blinov:2019rhb} are also presented.  Some existing or potential exclusion limits on conventional coupling $g_{a\gamma\gamma}$ are shown for reference.
}
\label{fig:ULDMsen}
\end{figure}

\begin{figure}[htb!]
\centering
\includegraphics[scale=0.13]{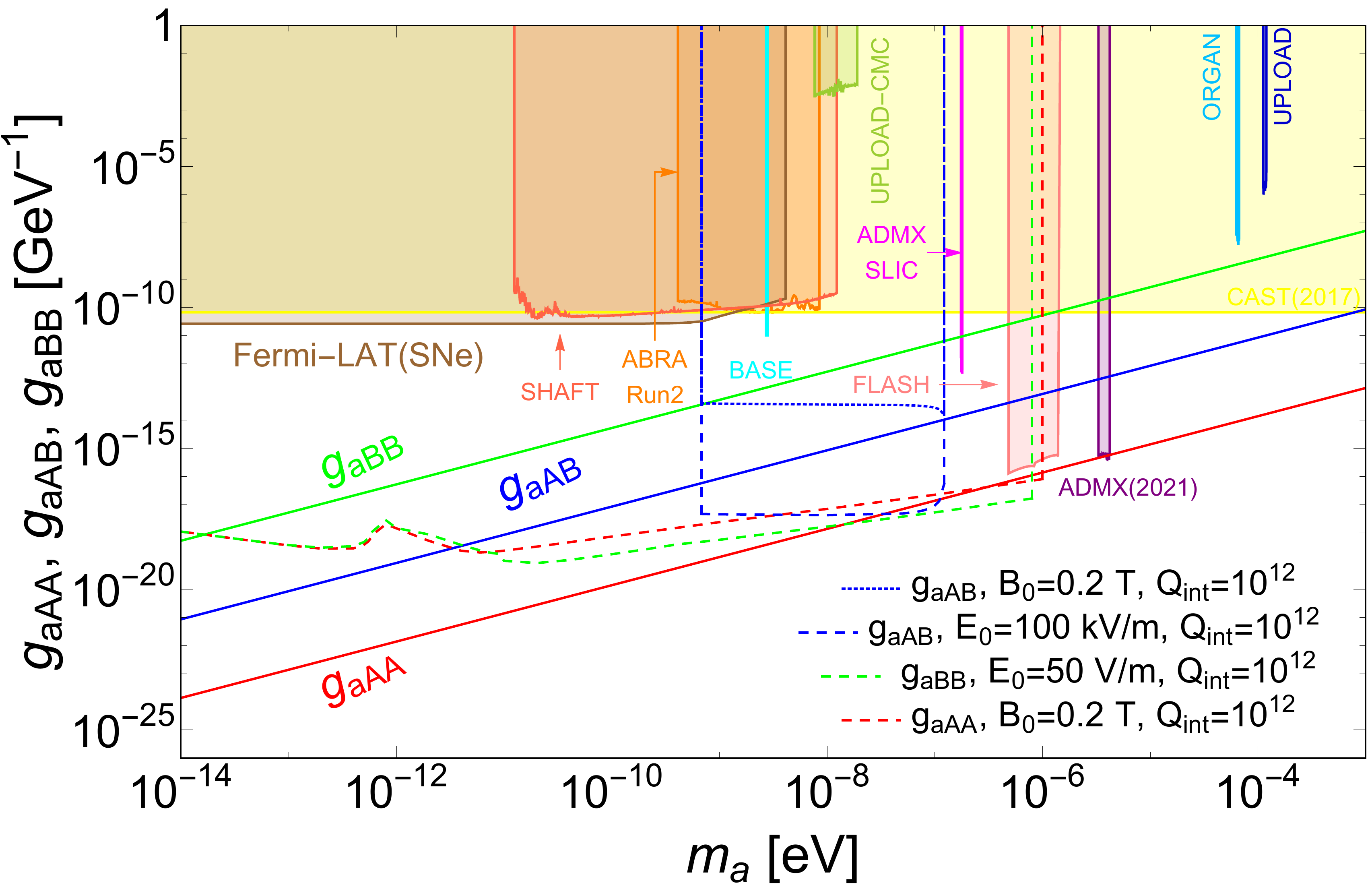}
\caption{
The expected sensitivity bounds of $g_{aAA}$ with $B_0=0.2~{\rm T}$ (red dashed line), $g_{aAB}$ with $B_0=0.2~{\rm T}$ (blue dotted line) and $E_0=100~{\rm kV/m}$ (blue dashed line) and $g_{aBB}$ with $E_0=50~{\rm kV/m}$ (green dashed line). The theoretical predictions of $g_{aAA}$ (red solid line), $g_{aAB}$ (blue solid line) and $g_{aBB}$ (green solid line)~\cite{Sokolov:2022fvs} are also presented. Some existing or potential exclusion limits are shown for reference.
}
\label{fig:QEMDsen}
\end{figure}

In Figs.~\ref{fig:ULDMsen} and \ref{fig:QEMDsen}, we show the expected sensitivity bounds on the generic couplings of low-mass DM in sub-$\mu$eV mass region. The bounds on conventional coupling $g_{a\gamma\gamma}$ and $g_{aAA}$ restore the result in Ref.~\cite{Berlin:2019ahk} with $Q_{\rm int}=10^{12}$. For $g_{aBB}$ coupling, we take a conservative magnitude of electric field $E_0=50~{\rm V/m}$. The axion mass lower than $10^{-6}$ eV can be probed by the axion-induced frequency conversion in SRF cavity with electric field. For $g_{a\gamma\gamma}^{\cancel{\rm CP}}$ and $g_{aAB}$ couplings, as discussed above, the reachable range of axion mass $10^{-9}\sim 10^{-7}$ eV is determined by the scope of adjusted cavity height $L$. To probe $g_{aAB}$ coupling, one needs $E_0\simeq 100~{\rm kV/m}$ which is lower than the breakdown electric field strength of a normal capacitor~\cite{Chen:2020cbs}.

%%%%%%%%%%%%%%%%%%%%%%

%%%%%%%%%%%%%%%%%%%%%%%%%%%%%%%%
\section{Conclusion}
\label{sec:Con}
%%%%%%%%%%%%%%%%%%%%%%%%%%%%%%%%%

In this work we investigate the detection of non-standard electromagnetic couplings in a few benchmark models of low-mass DM. For illustration, we consider the generic axion electrodynamics as well as the newly proposed axion electromagnetodynamics based on the QEMD framework.

The SRF cavity with two modes has more advantages than the traditional cavity approach with static background field or the LC circuit regime. It prefers a high quality factor and is sensitive to much lighter axion. We utilize the two-mode SRF cavity to probe the non-standard couplings of low-mass DM with frequency lower than GHz. We find that one can impose oscillating magnetic field and electric field to detect ordinary axion coupling $g_{a\gamma\gamma}$ and CP violating coupling $g_{a\gamma\gamma}^{\cancel{\rm CP}}$, respectively. In the QEMD axion framework, the background magnetic (electric) field can be introduced to probe both $g_{aAA}$ ($g_{aBB}$) and $g_{aAB}$ couplings.
The choices of the transverse electromagnetic modes and the index integers are limited by the corresponding selection rules. Finally, we show the sensitivity of the two-mode SRF cavity to the above non-standard axion couplings.

\acknowledgments
We would like to thank Michael E.~Tobar's helpful comments.
T.~L. is supported by the National Natural Science Foundation of China (Grant Nos. 12375096, 12035008, 11975129) and ``the Fundamental Research Funds for the Central Universities'', Nankai University (Grant No. 63196013).


\begin{thebibliography}{0}%
%\cite{Peccei:1977hh}
\bibitem{Peccei:1977hh}
R.~D.~Peccei and H.~R.~Quinn,
%``CP Conservation in the Presence of Instantons,''
Phys. Rev. Lett. \textbf{38}, 1440-1443 (1977)
doi:10.1103/PhysRevLett.38.1440
%7997 citations counted in INSPIRE as of 26 Aug 2024

%\cite{Peccei:1977ur}
\bibitem{Peccei:1977ur}
R.~D.~Peccei and H.~R.~Quinn,
%``Constraints Imposed by CP Conservation in the Presence of Instantons,''
Phys. Rev. D \textbf{16}, 1791-1797 (1977)
doi:10.1103/PhysRevD.16.1791
%4146 citations counted in INSPIRE as of 26 Aug 2024

%\cite{Weinberg:1977ma}
\bibitem{Weinberg:1977ma}
S.~Weinberg,
%``A New Light Boson?,''
Phys. Rev. Lett. \textbf{40}, 223-226 (1978)
doi:10.1103/PhysRevLett.40.223
%5681 citations counted in INSPIRE as of 26 Aug 2024

%\cite{Wilczek:1977pj}
\bibitem{Wilczek:1977pj}
F.~Wilczek,
%``Problem of Strong  $P$  and  $T$  Invariance in the Presence of Instantons,''
Phys. Rev. Lett. \textbf{40}, 279-282 (1978)
doi:10.1103/PhysRevLett.40.279
%5448 citations counted in INSPIRE as of 26 Aug 2024

%\cite{Baluni:1978rf}
\bibitem{Baluni:1978rf}
V.~Baluni,
%``CP Violating Effects in QCD,''
Phys. Rev. D \textbf{19}, 2227-2230 (1979)
doi:10.1103/PhysRevD.19.2227
%704 citations counted in INSPIRE as of 26 Aug 2024

%\cite{Crewther:1979pi}
\bibitem{Crewther:1979pi}
R.~J.~Crewther, P.~Di Vecchia, G.~Veneziano and E.~Witten,
%``Chiral Estimate of the Electric Dipole Moment of the Neutron in Quantum Chromodynamics,''
Phys. Lett. B \textbf{88}, 123 (1979)
[erratum: Phys. Lett. B \textbf{91}, 487 (1980)]
doi:10.1016/0370-2693(79)90128-X
%1062 citations counted in INSPIRE as of 26 Aug 2024

%\cite{Kim:1979if}
\bibitem{Kim:1979if}
J.~E.~Kim,
%``Weak Interaction Singlet and Strong CP Invariance,''
Phys. Rev. Lett. \textbf{43}, 103 (1979)
doi:10.1103/PhysRevLett.43.103
%3139 citations counted in INSPIRE as of 26 Aug 2024

%\cite{Shifman:1979if}
\bibitem{Shifman:1979if}
M.~A.~Shifman, A.~I.~Vainshtein and V.~I.~Zakharov,
%``Can Confinement Ensure Natural CP Invariance of Strong Interactions?,''
Nucl. Phys. B \textbf{166}, 493-506 (1980)
doi:10.1016/0550-3213(80)90209-6
%2812 citations counted in INSPIRE as of 26 Aug 2024

%\cite{Dine:1981rt}
\bibitem{Dine:1981rt}
M.~Dine, W.~Fischler and M.~Srednicki,
%``A Simple Solution to the Strong CP Problem with a Harmless Axion,''
Phys. Lett. B \textbf{104}, 199-202 (1981)
doi:10.1016/0370-2693(81)90590-6
%3553 citations counted in INSPIRE as of 26 Aug 2024

%\cite{Zhitnitsky:1980tq}
\bibitem{Zhitnitsky:1980tq}
A.~R.~Zhitnitsky,
%``On Possible Suppression of the Axion Hadron Interactions. (In Russian),''
Sov. J. Nucl. Phys. \textbf{31}, 260 (1980)
%2278 citations counted in INSPIRE as of 26 Aug 2024

%\cite{Baker:2006ts}
\bibitem{Baker:2006ts}
C.~A.~Baker, D.~D.~Doyle, P.~Geltenbort, K.~Green, M.~G.~D.~van der Grinten, P.~G.~Harris, P.~Iaydjiev, S.~N.~Ivanov, D.~J.~R.~May and J.~M.~Pendlebury, \textit{et al.}
%``An Improved experimental limit on the electric dipole moment of the neutron,''
Phys. Rev. Lett. \textbf{97}, 131801 (2006)
doi:10.1103/PhysRevLett.97.131801
[arXiv:hep-ex/0602020 [hep-ex]].
%1616 citations counted in INSPIRE as of 26 Aug 2024

%\cite{Pendlebury:2015lrz}
\bibitem{Pendlebury:2015lrz}
J.~M.~Pendlebury, S.~Afach, N.~J.~Ayres, C.~A.~Baker, G.~Ban, G.~Bison, K.~Bodek, M.~Burghoff, P.~Geltenbort and K.~Green, \textit{et al.}
%``Revised experimental upper limit on the electric dipole moment of the neutron,''
Phys. Rev. D \textbf{92}, no.9, 092003 (2015)
doi:10.1103/PhysRevD.92.092003
[arXiv:1509.04411 [hep-ex]].
%554 citations counted in INSPIRE as of 26 Aug 2024



%\cite{DiLuzio:2023cuk}
\bibitem{DiLuzio:2023cuk}
L.~Di Luzio, G.~Levati and P.~Paradisi,
%``The chiral Lagrangian of CP-violating axion-like particles,''
JHEP \textbf{2024}, no.02, 020 (2024)
doi:10.1007/JHEP02(2024)020
[arXiv:2311.12158 [hep-ph]].
%9 citations counted in INSPIRE as of 26 Aug 2024


%\cite{Luo:2023cxo}
\bibitem{Luo:2023cxo}
X.~Luo and A.~Mathur,
%``Cosmic birefringence from CP-violating axion interactions,''
JHEP \textbf{08}, 038 (2024)
doi:10.1007/JHEP08(2024)038
[arXiv:2311.03536 [hep-ph]].
%2 citations counted in INSPIRE as of 26 Aug 2024


%\cite{Gorghetto:2021luj}
\bibitem{Gorghetto:2021luj}
M.~Gorghetto, G.~Perez, I.~Savoray and Y.~Soreq,
%``Probing CP violation in photon self-interactions with cavities,''
JHEP \textbf{10}, 056 (2021)
doi:10.1007/JHEP10(2021)056
[arXiv:2103.06298 [hep-ph]].
%9 citations counted in INSPIRE as of 26 Aug 2024



%\cite{Schwinger:1966nj}
\bibitem{Schwinger:1966nj}
J.~S.~Schwinger,
%``Magnetic charge and quantum field theory,''
Phys. Rev. \textbf{144}, 1087-1093 (1966)
doi:10.1103/PhysRev.144.1087
%491 citations counted in INSPIRE as of 26 Aug 2024

%\cite{Zwanziger:1968rs}
\bibitem{Zwanziger:1968rs}
D.~Zwanziger,
%``Quantum field theory of particles with both electric and magnetic charges,''
Phys. Rev. \textbf{176}, 1489-1495 (1968)
doi:10.1103/PhysRev.176.1489
%290 citations counted in INSPIRE as of 26 Aug 2024

%\cite{Zwanziger:1970hk}
\bibitem{Zwanziger:1970hk}
D.~Zwanziger,
%``Local Lagrangian quantum field theory of electric and magnetic charges,''
Phys. Rev. D \textbf{3}, 880 (1971)
doi:10.1103/PhysRevD.3.880
%354 citations counted in INSPIRE as of 26 Aug 2024


%\cite{Sokolov:2022fvs}
\bibitem{Sokolov:2022fvs}
A.~V.~Sokolov and A.~Ringwald,
%``Electromagnetic Couplings of Axions,''
[arXiv:2205.02605 [hep-ph]].
%36 citations counted in INSPIRE as of 26 Aug 2024

%\cite{Sokolov:2023pos}
\bibitem{Sokolov:2023pos}
A.~V.~Sokolov and A.~Ringwald,
%``Generic Axion Maxwell Equations: Path Integral Approach,''
Annalen Phys. \textbf{2023}, 2300112 (2023)
doi:10.1002/andp.202300112
[arXiv:2303.10170 [hep-ph]].
%14 citations counted in INSPIRE as of 26 Aug 2024

%\cite{Heidenreich:2023pbi}
\bibitem{Heidenreich:2023pbi}
B.~Heidenreich, J.~McNamara and M.~Reece,
%``Non-standard axion electrodynamics and the dual Witten effect,''
JHEP \textbf{01}, 120 (2024)
doi:10.1007/JHEP01(2024)120
[arXiv:2309.07951 [hep-ph]].
%13 citations counted in INSPIRE as of 26 Aug 2024

%\cite{Li:2023zcp}
\bibitem{Li:2023zcp}
T.~Li and R.~J.~Zhang,
%``The electroweak magnetic monopole in the presence of KSVZ axion,''
[arXiv:2312.01355 [hep-ph]].
%2 citations counted in INSPIRE as of 26 Aug 2024

%\cite{Li:2022oel}
\bibitem{Li:2022oel}
T.~Li, R.~J.~Zhang and C.~J.~Dai,
%``Solutions to axion electromagnetodynamics and new search strategies of sub-\ensuremath{\mu}eV axion,''
JHEP \textbf{03}, 088 (2023)
doi:10.1007/JHEP03(2023)088
[arXiv:2211.06847 [hep-ph]].
%15 citations counted in INSPIRE as of 26 Aug 2024

%\cite{Tobar:2022rko}
\bibitem{Tobar:2022rko}
M.~E.~Tobar, C.~A.~Thomson, B.~T.~McAllister, M.~Goryachev, A.~V.~Sokolov and A.~Ringwald,
%``Sensitivity of Resonant Axion Haloscopes to Quantum Electromagnetodynamics,''
Annalen Phys. \textbf{536}, no.1, 2200594 (2024)
doi:10.1002/andp.202200594
[arXiv:2211.09637 [hep-ph]].
%16 citations counted in INSPIRE as of 26 Aug 2024

%\cite{McAllister:2022ibe}
\bibitem{McAllister:2022ibe}
B.~T.~McAllister, A.~Quiskamp, C.~A.~J.~O'Hare, P.~Altin, E.~N.~Ivanov, M.~Goryachev and M.~E.~Tobar,
%``Limits on Dark Photons, Scalars, and Axion-Electromagnetodynamics with the ORGAN Experiment,''
Annalen Phys. \textbf{536}, no.1, 2200622 (2024)
doi:10.1002/andp.202200622
[arXiv:2212.01971 [hep-ph]].
%18 citations counted in INSPIRE as of 26 Aug 2024

%\cite{Li:2023kfh}
\bibitem{Li:2023kfh}
T.~Li, C.~J.~Dai and R.~J.~Zhang,
%``Searching for high-frequency axion in quantum electromagnetodynamics through interface haloscopes,''
Phys. Rev. D \textbf{109}, no.1, 015026 (2024)
doi:10.1103/PhysRevD.109.015026
[arXiv:2304.12525 [hep-ph]].
%7 citations counted in INSPIRE as of 26 Aug 2024

%\cite{Li:2023aow}
\bibitem{Li:2023aow}
T.~Li and R.~J.~Zhang,
%``Quantum calculation of axion-photon transition in electromagnetodynamics for cavity haloscope*,''
Chin. Phys. C \textbf{47}, no.12, 123104 (2023)
doi:10.1088/1674-1137/ad0620
[arXiv:2305.01344 [hep-ph]].
%6 citations counted in INSPIRE as of 26 Aug 2024

%\cite{Tobar:2023rga}
\bibitem{Tobar:2023rga}
M.~E.~Tobar, A.~V.~Sokolov, A.~Ringwald and M.~Goryachev,
%``Searching for GUT-scale QCD axions and monopoles with a high-voltage capacitor,''
Phys. Rev. D \textbf{108}, no.3, 035024 (2023)
doi:10.1103/PhysRevD.108.035024
[arXiv:2306.13320 [hep-ph]].
%11 citations counted in INSPIRE as of 26 Aug 2024

%\cite{Patkos:2023lof}
\bibitem{Patkos:2023lof}
A.~Patkos,
%``Electromagnetic energy transfer processes in effective electro-magneto dynamics of axions,''
Mod. Phys. Lett. A \textbf{38}, no.30n31, 2350137 (2023)
doi:10.1142/S0217732323501377
[arXiv:2309.05523 [hep-ph]].
%4 citations counted in INSPIRE as of 26 Aug 2024

%\cite{Witten:1979ey}
\bibitem{Witten:1979ey}
E.~Witten,
%``Dyons of Charge e theta/2 pi,''
Phys. Lett. B \textbf{86}, 283-287 (1979)
doi:10.1016/0370-2693(79)90838-4
%922 citations counted in INSPIRE as of 18 Oct 2024

%\cite{Sikivie:1983ip}
\bibitem{Sikivie:1983ip}
P.~Sikivie,
%``Experimental Tests of the Invisible Axion,''
Phys. Rev. Lett. \textbf{51}, 1415-1417 (1983)
[erratum: Phys. Rev. Lett. \textbf{52}, 695 (1984)]
doi:10.1103/PhysRevLett.51.1415
%2028 citations counted in INSPIRE as of 26 Aug 2024

%\cite{Sikivie:2010fa}
\bibitem{Sikivie:2010fa}
P.~Sikivie,
%``Superconducting Radio Frequency Cavities as Axion Dark Matter Detectors,''
[arXiv:1009.0762 [hep-ph]].
%23 citations counted in INSPIRE as of 26 Aug 2024

%\cite{Goryachev:2018vjt}
\bibitem{Goryachev:2018vjt}
M.~Goryachev, B.~Mcallister and M.~E.~Tobar,
%``Axion detection with precision frequency metrology,''
Phys. Dark Univ. \textbf{26}, 100345 (2019)
[erratum: Phys. Dark Univ. \textbf{32}, 100787 (2021)]
doi:10.1016/j.dark.2019.100345
[arXiv:1806.07141 [physics.ins-det]].
%37 citations counted in INSPIRE as of 26 Aug 2024

%\cite{Thomson:2019aht}
\bibitem{Thomson:2019aht}
C.~A.~Thomson, B.~T.~McAllister, M.~Goryachev, E.~N.~Ivanov and M.~E.~Tobar,
%``Upconversion Loop Oscillator Axion Detection Experiment: A Precision Frequency Interferometric Axion Dark Matter Search with a Cylindrical Microwave Cavity,''
Phys. Rev. Lett. \textbf{126}, no.8, 081803 (2021)
[erratum: Phys. Rev. Lett. \textbf{127}, no.1, 019901 (2021)]
doi:10.1103/PhysRevLett.127.019901
[arXiv:1912.07751 [hep-ex]].
%41 citations counted in INSPIRE as of 26 Aug 2024

%\cite{Berlin:2019ahk}
\bibitem{Berlin:2019ahk}
A.~Berlin, R.~T.~D'Agnolo, S.~A.~R.~Ellis, C.~Nantista, J.~Neilson, P.~Schuster, S.~Tantawi, N.~Toro and K.~Zhou,
%``Axion Dark Matter Detection by Superconducting Resonant Frequency Conversion,''
JHEP \textbf{07}, no.07, 088 (2020)
doi:10.1007/JHEP07(2020)088
[arXiv:1912.11048 [hep-ph]].
%71 citations counted in INSPIRE as of 26 Aug 2024

%\cite{Lasenby:2019prg}
\bibitem{Lasenby:2019prg}
R.~Lasenby,
%``Microwave cavity searches for low-frequency axion dark matter,''
Phys. Rev. D \textbf{102}, no.1, 015008 (2020)
doi:10.1103/PhysRevD.102.015008
[arXiv:1912.11056 [hep-ph]].
%40 citations counted in INSPIRE as of 26 Aug 2024

%\cite{Lasenby:2019hfz}
\bibitem{Lasenby:2019hfz}
R.~Lasenby,
%``Parametrics of Electromagnetic Searches for Axion Dark Matter,''
Phys. Rev. D \textbf{103}, no.7, 075007 (2021)
doi:10.1103/PhysRevD.103.075007
[arXiv:1912.11467 [hep-ph]].
%23 citations counted in INSPIRE as of 26 Aug 2024

%\cite{Bogorad:2019pbu}
\bibitem{Bogorad:2019pbu}
Z.~Bogorad, A.~Hook, Y.~Kahn and Y.~Soreq,
%``Probing Axionlike Particles and the Axiverse with Superconducting Radio-Frequency Cavities,''
Phys. Rev. Lett. \textbf{123}, no.2, 021801 (2019)
doi:10.1103/PhysRevLett.123.021801
[arXiv:1902.01418 [hep-ph]].
%55 citations counted in INSPIRE as of 26 Aug 2024

%\cite{Berlin:2020vrk}
\bibitem{Berlin:2020vrk}
A.~Berlin, R.~T.~D'Agnolo, S.~A.~R.~Ellis and K.~Zhou,
%``Heterodyne broadband detection of axion dark matter,''
Phys. Rev. D \textbf{104}, no.11, L111701 (2021)
doi:10.1103/PhysRevD.104.L111701
[arXiv:2007.15656 [hep-ph]].
%73 citations counted in INSPIRE as of 26 Aug 2024

%\cite{Gao:2020anb}
\bibitem{Gao:2020anb}
C.~Gao and R.~Harnik,
%``Axion searches with two superconducting radio-frequency cavities,''
JHEP \textbf{07}, 053 (2021)
doi:10.1007/JHEP07(2021)053
[arXiv:2011.01350 [hep-ph]].
%22 citations counted in INSPIRE as of 26 Aug 2024

%\cite{Salnikov:2020urr}
\bibitem{Salnikov:2020urr}
D.~Salnikov, P.~Satunin, D.~V.~Kirpichnikov and M.~Fitkevich,
%``Examining axion-like particles with superconducting radio-frequency cavity,''
JHEP \textbf{03}, 143 (2021)
doi:10.1007/jhep03(2021)143
[arXiv:2011.12871 [hep-ph]].
%12 citations counted in INSPIRE as of 26 Aug 2024

%\cite{Berlin:2022hfx}
\bibitem{Berlin:2022hfx}
A.~Berlin, S.~Belomestnykh, D.~Blas, D.~Frolov, A.~J.~Brady, C.~Braggio, M.~Carena, R.~Cervantes, M.~Checchin and C.~Contreras-Martinez, \textit{et al.}
%``Searches for New Particles, Dark Matter, and Gravitational Waves with SRF Cavities,''
[arXiv:2203.12714 [hep-ph]].
%41 citations counted in INSPIRE as of 26 Aug 2024

%% This BibTeX bibliography file was created using BibDesk.
%% http://bibdesk.sourceforge.net/

%\cite{Giaccone:2022pke}
%\bibitem{Giaccone:2022pke}
%B.~Giaccone, A.~Berlin, I.~Gonin, A.~Grassellino, R.~Harnik, Y.~Kahn, T.~Khabiboulline, A.~Lunin, A.~Melnychuk and A.~Netepenko, \textit{et al.}
%``Design of axion and axion dark matter searches based on ultra high Q SRF cavities,''
%[arXiv:2207.11346 [hep-ex]].
%6 citations counted in INSPIRE as of 26 Aug 2024

%\cite{Romanenko:2023irv}
%\bibitem{Romanenko:2023irv}
%A.~Romanenko, R.~Harnik, A.~Grassellino, R.~Pilipenko, Y.~Pischalnikov, Z.~Liu, O.~S.~Melnychuk, B.~Giaccone, O.~Pronitchev and T.~Khabiboulline, \textit{et al.}
%``Search for Dark Photons with Superconducting Radio Frequency Cavities,''
%Phys. Rev. Lett. \textbf{130}, no.26, 261801 (2023)
%doi:10.1103/PhysRevLett.130.261801
%[arXiv:2301.11512 [hep-ex]].
%32 citations counted in INSPIRE as of 26 Aug 2024

%\cite{SHANHE:2023kxz}
%\bibitem{SHANHE:2023kxz}
%Z.~Tang \textit{et al.} [SHANHE],
%``First Scan Search for Dark Photon Dark Matter with a Tunable Superconducting Radio-Frequency Cavity,''
%Phys. Rev. Lett. \textbf{133}, no.2, 021005 (2024)
%doi:10.1103/PhysRevLett.133.021005
%[arXiv:2305.09711 [hep-ex]].
%12 citations counted in INSPIRE as of 26 Aug 2024

%\cite{ParticleDataGroup:2022pth}
%\bibitem{ParticleDataGroup:2022pth}
%R.~L.~Workman \textit{et al.} [Particle Data Group],
%``Review of Particle Physics,''
%PTEP \textbf{2022}, 083C01 (2022)
%doi:10.1093/ptep/ptac097
%3729 citations counted in INSPIRE as of 26 Aug 2024

%\cite{Bernard:2001kp}
\bibitem{Bernard:2001kp}
P.~Bernard, G.~Gemme, R.~Parodi and E.~Picasso,
%``A Detector of small harmonic displacements based on two coupled microwave cavities,''
Rev. Sci. Instrum. \textbf{72}, 2428-2437 (2001)
doi:10.1063/1.1366636
[arXiv:gr-qc/0103006 [gr-qc]].
%53 citations counted in INSPIRE as of 26 Aug 2024

%\cite{Chen:2020cbs}
\bibitem{Chen:2020cbs}
W.~Chen, Y.~Gao and Q.~Yang,
%``Broadband dark matter axion detection using a cylindrical capacitor,''
Nucl. Phys. B \textbf{1005}, 116602 (2024)
doi:10.1016/j.nuclphysb.2024.116602
[arXiv:2012.13946 [hep-ph]].
%3 citations counted in INSPIRE as of 26 Aug 2024

%\cite{DiLuzio:2020wdo}
\bibitem{DiLuzio:2020wdo}
L.~Di Luzio, M.~Giannotti, E.~Nardi and L.~Visinelli,
%``The landscape of QCD axion models,''
Phys. Rept. \textbf{870}, 1-117 (2020)
doi:10.1016/j.physrep.2020.06.002
[arXiv:2003.01100 [hep-ph]].
%677 citations counted in INSPIRE as of 26 Aug 2024

%\cite{Blinov:2019rhb}
\bibitem{Blinov:2019rhb}
N.~Blinov, M.~J.~Dolan, P.~Draper and J.~Kozaczuk,
%``Dark matter targets for axionlike particle searches,''
Phys. Rev. D \textbf{100}, no.1, 015049 (2019)
doi:10.1103/PhysRevD.100.015049
[arXiv:1905.06952 [hep-ph]].
%47 citations counted in INSPIRE as of 26 Aug 2024


\end{thebibliography}
\end{document}